# A Social Network Framework to Explore Healthcare Collaboration

**Uma Srinivasan**
Lead Scientist, Capital Markets Cooperative Research Centre
55 Harrington Street, Sydney 2000, Australia
Phone: +61 02 8088 4234
Fax: +61 02 8088 4201
E-mail: umas@cmcrc.com

**\*Shahadat Uddin**
Lecturer, Project Management Program & Centre for Complex Systems
The University of Sydney, Australia
Phone: +61 02 9351 2118
Fax: +61 02 9662 6566
Email: shahadat.uddin@sydney.edu.au

\* Corresponding Author

**ACKNOWLEDGEMENT**

*The authors would like to thank the anonymous reviewers and the editor for their insightful comments and suggestions.*

# Biographical Notes

**Dr Uma Srinivasan** is a Lead Scientist at Capital Markets Cooperative Research Centre and CMC Health Insurance Solutions, Sydney. Her experience in designing software solutions for hospitals and health departments has shaped her research focus on technologies and solutions that enable efficient and high quality healthcare. She has several publications in leading international books, journals and conference proceedings. Uma holds a PhD in computer science from University of New South Wales, Australia.

**Dr Shahadat Uddin** conducts research in the area of complex healthcare coordination and collaboration networks, health informatics, and data analysis and modelling. His research addresses interdisciplinary issues to understand the impact of network structure and network dynamics on group performance and coordination outcomes, in complex, dynamic and distributed environments. He holds a PhD in network science from the University of Sydney.

# A Social Network Framework to Explore Healthcare Collaboration

**ABSTRACT**

A patient centric approach to healthcare calls for increased collaboration among healthcare professionals who look after patients, both in and out of the hospital. As a result an informal social network emerges among healthcare professionals who collaborate while looking after patients. The nature of such a collaboration network whose purpose is to address a specific health problem raises some interesting questions. Using principles of social network theories (i.e. Bavelas' Centrality Theory (1950) and Freeman's centralization theory (1978)) and the social network model of exponential random graph model, this paper presents a research framework to explore and analyse the collaboration network that evolves among healthcare professionals during the course of treating patients. The proposed research framework: (i) identifies the type of collaboration structure among physicians that is effective and efficient for patients, in terms of outcome measures such as hospitalisation cost and length of stay; (ii) explores and identifies the effective structural attributes of a referral network that evolves during the course of providing care; and (iii) explores the impact of socio-demographic characteristics of healthcare professionals, patients and hospitals on collaboration structures from the point of view of measurable outcomes such as cost and quality of care.

The framework presented in this paper uses illustrative examples of collaboration that evolves among physicians while treating hospital patients undergoing hip replacement surgery and subsequent rehabilitation process. This practical application of the proposed framework reveals structures of physicians' collaborations that are not favourable to cost and quality of care measure such as readmission rate. We believe that such a framework will enable healthcare managers and administrators to evaluate the collaborative work environment within their respective healthcare organisations.

# A Social Network Framework to Explore Healthcare Collaboration

Uma Srinivasan, Capital Markets Cooperative Research Centre, Sydney
Shahadat Uddin, University of Sydney, Australia

## INTRODUCTION

Healthcare spending is a major topic of discussion in practically every country in the world. Figure 1 shows the health spending as a percentage of GDP of a few Organisation for Economic Co-operation and Development (OECD) countries based on a report produced by the International Federation of Health Plans[1]. There are growing concerns all around about spiralling healthcare costs, budget constraints and their impact on quality of health outcomes for patients. In this context, a major goal of governments and health providers is to achieve consistency of health outcomes for frequent and expensive health services and high volume and high cost procedures such as knee and hip replacement surgeries whose numbers have been steadily increasing over the years.

(Introduce Figure 1 here)

Figure 2 shows the comparative costs of hip replacement procedures from the same report. Comparisons across different countries are complicated by differences in sectors, fee schedules and health plans used for cost comparisons. Nevertheless, it is clear that consistent healthcare strategies are required to deliver high quality of services where health outcomes are consistent and predictable both for the patient and providers involved in caring for the patient.

(Introduce Figure 2 here)

Although not always formal, there is a protocol among physicians to collaborate while providing care to chronic patients both in a hospital setting as well as in an ambulatory care setting. Given that an informal social network exists or emerges among healthcare professionals to address a specific problem, the question that arises is whether there is a way to measure the network parameters of collaboration networks that consistently perform effectively and efficiently to achieve the desired outcome which could be, in the case of patient care, high quality care with optimal costs. Using principles of social network theories such as Bavelas' Centrality Theory (1950), Freeman's centralization theory (1978), and the social network model of exponential random graph model, we propose a social network-based framework that can serve as a basis to offer insights into the different types of collaboration patterns among healthcare professionals that are conducive to positive health outcomes to patients, as well as provide consistent quality of care measures in healthcare settings.

---

[1] http://hushp.harvard.edu/sites/default/files

The rest of the paper is organised as follows: the next section presents a review of network-based collaboration models in a healthcare setting, which is followed by the theoretical background of healthcare collaboration related to network structures and measures of centrality. Then we presents our network-based framework that provides formal measures of physicians' collaboration networks and identifies network measures that indicate positive outcomes in terms of both costs and quality of care. After that, we presents an illustrative application and results of network structure measures on a specific (de-identified and anonymised) health insurance claims data set provided by an Australian not-for-profit health insurance organisation. The last section discusses the contribution of this paper, and presents conclusions and future research directions.

## REVIEW OF HEALTHCARE COLLABORATION

There are numerous studies in current literature exploring collaborations among healthcare professionals. Most of these studies examine hospital performance and patient outcomes by analysing collaboration networks among different healthcare professionals such as nurse-physician collaboration (Knaus et al., 1986), physician-pharmacist collaboration (Hunt et al., 2008), physician-patient collaboration (Arbuthnott & Sharpe, 2009), hospital-physician collaboration (Burns & Muller, 2008), and inter-professional and interdisciplinary collaboration (Gaboury et al., 2009). Cunningham et al. (2012) have conducted an orderly review of 26 studies of professionals' network structures and analysed factors connected with network effectiveness and sustainability specifically in relation to the quality of care and patient safety. They noticed that cohesive and collaborative health professional networks can contribute to improving quality and safety of care.

Another classic study, led by Knaus and his team, identifies a significant relationship between the degree of nurse-physician collaboration and patient mortality in intensive care units (Knaus, et al., 1986). They study treatment and outcome in 5030 intensive care unit patients and find that hospitals where nurse-physician collaboration is present report a lower mortality rate compared to the predicted number of patient deaths. Conversely, hospitals that are noted for poor communication among healthcare professionals exceed their predicted number of patient deaths. In a two group quasi-experiment on 1207 general medicine patients (n = 581 in the experimental group who received care from a specially designed care management plan that facilitated higher collaboration among hospital staff and n = 626 in the control group who received the usual care), Cowan et al. (2006) notice average hospital length of stay, total hospitalization cost and hospital readmission rate are significantly lower for patients in the experimental group than the control group (5 versus 6 days, p< .0001) which contributes a *'backfill profit'* of US$1591 per patient to hospitals. Sommers et al. (2000) examined the impact of an interdisciplinary and collaborative practice intervention involving a primary care physician, a nurse and a social worker for community-dwelling seniors with chronic illnesses. They conducted a controlled cohort study of 543 patients in 18 private office practices of primary care physicians. The intervention group received care from their primary care physician working with a registered nurse and a social worker, while the control group received care as usual from primary care physicians. They noticed that the intervention group produced better results in relation to readmission rates and average office visits to all physicians. Moreover, the patients in the intervention group also reported an increase in social activities compared with the control group.

There are other studies emphasizing collaboration for effective patient outcome across professional boundaries within hospitals. By analysing data collected from 105 interviews (with 40 physician, 32 case managers, 23 physician office staff, 8 administrators and 2 case assistants), Netting and Williams (1996) argue that there is a growing need to collaborate and communicate across professional lines rather than make assumptions about who can do what for better patient outcomes, professional satisfaction and hospital performance. There are other studies that analyse networked collaboration among healthcare specialists to explore different aspects of professional behaviour and quality patient care. For example, Fattore et al. (2009) evaluate the effects of GP network organisation on their prescribing behavior and (Meltzer et al., 2010) develop a selection criteria of group members in order to improve the effectiveness of team-based approach to patient care.

# THEORETICAL MOTIVATION FOR COLLABORATION

Collaboration evolves among healthcare professionals during the course of providing healthcare services to patients. As a result of this collaboration an informal social network emerges among healthcare professionals over time. A social network is viewed as a set of actors and a set of links between those actors. In a social network, an actor is a node which represents an entity such as an individual or an organisation. The formation of a social network is typically associated with the need for an actor to receive some sort of information or resource from others. Each actor is a part of this informal network although they have different positions and locations in the network. Since all actors do not have the same network locations, a reasonable question arises: *"Do actors, who have different network locations, show notable differences in their ability to collaborate?"* A review of classical network theories (i.e. Bavelas Experiment and Freeman's Theory of Centralisation) can provide a better understanding to respond to this performance-related question. These two network theories explain structural influences of collaboration network on group performance.

## Bavelas's experiment

The *"Bavelas Experiment"* was conducted by Alex Bavelas and his students, particularly Harold Leavitt, in the Small Group Network Laboratory at *Massachusetts Institute of Technology* (MIT) in the late 1940s (Bavelas, 1950; H. Leavitt, 1949). This experiment, which is also known as the MIT experiment, attempted to seek the impact of different types of human communication patterns on performance. Based on the assumption that (i) success of entire classes of tasks depends upon an effective flow of information, and (ii) communication patterns have effects on task performance and individual outcome, this study focused on the motivating question of *"On what principles may a pattern of communication be determined that will in fact be a fit one for effective and efficient human effort?"* On the basis of the assumption that imposed patterns determine certain aspects of group process, in addition to the first two assumptions, the original driving question further raised three subsequent questions: *"How does a fixed communication pattern affect the performance of a group?"*; *"Do some patterns have structural properties that limit group performance?"*; and *"What effects can pattern have upon the emergence of leadership, the development of organisation, the degree of resistance to group disruption, the ability to adapt successfully to sudden changes in the working environment?"* These research questions sought to answer, through a laboratory-controlled experiment, how social network structures, measured in terms of patterns of communication, affect individual performance, and the work and life of a group (e.g. emergence of leaders and job satisfaction).

The experiment consisted of five people or subjects who had to communicate with each other only through enclosed cubicles in order to solve a puzzle. Each subject was given a card with five different symbols had been printed on it: a circle, a triangle, an asterisk, a plus sign and a diamond (O, Δ, *, +, ◊). Each symbol appeared at most four times in a group of five cards although only one symbol appeared on all five cards. The puzzle was to find the common symbol in the shortest time possible and with minimum communication among subjects. The cubicle of each subject had six switches labelled with one of the six master symbols. The puzzle was considered solved when each subject indicated the common symbol by throwing the appropriate switch. The experiment was trialled 15 times. The same six master symbols were used throughout, however; the common symbol varied from trial to trial. The subjects communicated by writing messages which were passed through slots in the walls of the cubicles. These slots were so arranged that any of the desired patterns of communication structures (i.e. the Y, star, circle and line) shown in Figure 3 could emerge.

[Insert Figure 3 here]

The subjects could pass as many messages as they wished and no restriction was placed on the content of these messages. None of the subjects had any idea about the number of people in the study and the configuration of the group communication structure. Total number of individuals and group times, and number of errors (i.e. when a subject threw an incorrect switch) in solving the puzzle were collected and recorded by a laboratory assistant for analysis purpose.

The performance and behaviour of all communication patterns were evaluated on the basis of structure comparison and node-level analysis. Performance of the task-oriented groups was measured in terms of the time taken to solve the puzzle and the number of errors made by each group in the course of completing the puzzle. When completion time (e.g. time needed to complete the puzzle) was considered to compare the performance of groups using various patterns, it turned out that the groups using the *'star'* and *'Y'* structures took less time on average than the groups using other structures (i.e. *'circle'* and *'line'*). Centralisation, as explained by Leavitt (1951), is the chief determinant to this performance difference for various communication structures. It was evident that patterns demonstrating higher centralisation performed better. When there was a central subject present in the structure, all the remaining subjects in the structure channelled all required information through that central subject. Therefore, the groups using *'star'* and *'Y'* structures also used fewer numbers of messages compared to groups using other structures and also made the least number of errors.

When node-level analysis was conducted to understand the behavioural differences of communication patterns, they discovered that highly centralised structures had a leader who emerged during the task process. In all structures, leaders emerged at those positions with the highest centrality. As the *'star'* and *'Y'* structures clearly had only one subject with highest degree-centrality, there were more chances of a leader evolving during the experiment, which in turn led to better performance. They also noted that subjects' satisfaction with their roles in the course of completing the puzzle varied among structures as well as among different positions within the same structure (i.e. central position versus peripheral position).

The key finding from this study was that centralised structures such as the *'star'* or *'Y'* were far more conducive to performance (i.e. solving the puzzle faster) than the decentralised or flatter structures such as the *'circle'* structure. The basic logic is that, in decentralised networks the information floats around inefficiently, and is therefore less conducive to efficient performance.

**Freeman's theory of centrality and centralisation**

The idea of centrality was applied to human communication by Bavelas (1950) in the late 1950s as described in the previous section. All experiments conducted by Bavelas and his research team concluded that centrality was related to group efficiency in problem-solving, perception of leadership and the personal satisfaction of participants. In the late 1970s, Freeman (1978) wrote a seminal article about the intuitive background for measures of structural centrality which immediately became one of the core concepts in the social network study. His work extended the notion of centrality by proposing three measures: (i) one absolute, (ii) one relative, and (iii) centralisation of the entire network. Each of these has three distinctive concepts for centrality: (a) degree, (b) betweenness, and (c) closeness. He also examined the implications of these nine measures (each of three centrality concepts has three measures) to a study of small groups.

Freeman (1978), motivated by the notion of centrality in Bevalas experiment, reviewed other studies that demonstrated the applicability of the concept of centrality to understand (i) real life problems such as political integration in the context of the diversity of Indian social life (Cohn & Marriott, 1958), (ii) consequences of centrality in communication paths for urban development (Pitts, 1965), and (iii) implications of centrality for the design of organisations (Beauchamp, 1965). Freeman (1978) then reviewed the various measures and overlapping concepts of centrality, while clarifying range, limitations and strength of each measure for application. Centrality had been defined in terms of *'point'*, *'betweenness'* and *'closeness'* as reported by Freeman (1978), while each of them has important implications on social outcomes, process and performance. Point centrality can be measured in terms of degree which is the number of ties to and from an actor in a network. Structurally, centrality is measured in terms of closeness and betweenness. Closeness centrality indicates the extent to which an actor is close to all others in the network, and betweenness centrality reflects the extent to which an actor lies in the shortest path to all others in the network. Each of the centrality concepts has been related to important social occurrences: *'point'* or *'degree'* centrality being viewed as an indicator of the communication activity of an actor; *'betweenness'* centrality being viewed as an indicator of the potential of communication control capacity of an actor; and *'closeness'* centrality is perceived as an index of minimum cost of time and efficiency to communicate with others in the network.

In a subsequent study, Freeman et al. (1979) explored the effects of structural centrality on human communication through the use of a replication of the early MIT experiments by Bavelas (1950). It was shown that although two of the three concepts of centrality measures had a demonstrable effect on individual responses and group processes, the classic measure of centrality based on distance was found unrelated to any experimental variables. Using 100 volunteers from among the student body at Leigh University as subjects, Freeman et al. (1979) analysed the results and demonstrated that centrality is an important structural factor of any network which influences leadership evolvement, and satisfaction and efficiency of actors or subjects. Interestingly, in their research another structural factor, the overall density of communication paths in the structural form, also turned out to be relevant in understanding network performance. Since then, the notion of centrality, density and centralisation were considered as one of the key network measures used for studying network effects on individual and group outcomes such as task efficiency, productivity, and improved performance (Alireza Abbasi et al., 2012; Ahuja et al., 2003; Bonacich, 1991; Brass, 1981, 1985; Cross & Cummings, 2004; Faust, 1997; Hossain et al., 2006; Mullen & Eduardo, 1991; Pfeffer, 1980; Salancik & Pfeffer, 1978; Sparrowe et al., 2001; M. Uddin & L. Hossain, 2009; M. S. Uddin & L. Hossain, 2009; M. S. Uddin & L. Hossain, 2011; Shahadat Uddin, Jafar Hamra, et al., 2013b; S. Uddin & L. Hossain, 2011; Uddin & Hossain, 2012; Uddin & Hossain, 2013; S. Uddin, L. Hossain, & M. Kelaher, 2012). Freeman's work made a substantial contribution to the network structure and task-performance research. Indeed, his contribution was so influential that the notion of centrality is now almost always attributed to him.

In summary, both Bavelas experiment and Freeman's Theory of Centrality and Centralisation open a new area of research to understand and explore individual performance in a collaborative environment. Bavelas (1950) divulged a new research area for investigating how network positions of actors influence their (i) ability to perform, (ii) perception of leadership, and (iii) level of personal satisfaction in the course of working in a collaborative environment towards achieving a common goal. Freeman (1978) proposed three measures for structural centrality: (i) degree centrality – indicating activity of actor and actor popularity, (ii) betweenness centrality – representing actor potential to control, and (iii) closeness centrality – stating the minimum cost to visit all other actors in the network. Since then, these three measures have been utilised extensively by researchers to measure structural positions of actors in a collaborative environment (A. Abbasi et al., 2011; Alireza Abbasi, et al., 2012; Cainelli et al., 2010; Shahian et al., 2010; Shippy et al., 2004; Smith et al., 2003; M. S. Uddin & L. Hossain, 2009; Shahadat Uddin et al., 2012; Uddin, et al., 2013b; S. Uddin, L. Hossain, A. Abbasi, et al., 2012; Shahadat Uddin, Liaquat Hossain, & Kim Rasmussen, 2013; Uddin & Jacobson, 2013; Uddin et al., 2011; S. Uddin et al., 2013; Uddin et al., 2014).

**PROPOSED COLLABORATION FRAMEWORK**

The proposed collaboration framework is based on the two network theories of centralisation and centrality, positioned in the context of collaboration that occurs among healthcare professionals while treating a patient with a specific problem. In particular, if the health problem is acute or chronic and requires hospitalisation, the patient goes through a complex journey from one provider to another, while negotiating through the maze of the health system. In the broader sense, most countries around the world have some versions of a '*patient journey*' that come into play when a patient needs hospital treatment. As an illustrative example, we briefly describe the different stages of the patient journey in the Australian healthcare setting.

All Australians are entitled to the government funded public healthcare, which is accessed through the Medicare system. In addition, most people have private health insurance to avoid waiting lists and have access to private hospital cover. Outpatient consultation costs are covered by Medicare and hospital treatment costs are borne by the private health insurance providers. Therefore, the journey involves negotiating both the public and private healthcare settings. Figure 4 shows a patient's journey through the healthcare system.

[Insert Figure 4 here]

Each box shows a service provider who provides a specific type of service to the patient. The entry point - the first point of contact - for all Australians is the General Practitioner (GP). Depending on the nature of the presenting problem, the GP could perform any one or all of the following activities: (i) give a script which is dispensed by the pharmacy; (ii) refer the patient for laboratory investigations; (iii) refer the patient for radiology and imaging services; (iv) refer the patient to one or more specialists; (v) send the patient to the public hospital in case of an emergency. All the above services including specialist consultations and emergency public hospital admissions are funded by the public Medicare system. When a patient requires hospitalisation for either investigative or therapeutic procedures, the specialist refers the patient for a hospital admission. At this point the journey changes its direction in different ways for public and private patients. Public patients who do not have private health cover will seek admission in a public hospital which may have long waiting periods for admission. Patients with private health insurance have the option of seeking admission in a private hospital or as a private patient in a public hospital or as public patient in a public hospital. The advantage of seeking admission as a private patient is the shorter waiting time for hospital admissions. For patients with private health insurance all *in-hospital* charges including surgeons' fees, multiple specialist fees, and laboratory and radiology charges are paid for by the private health insurer. Depending on the nature of the illness and treatment, the patient might be discharged and the journey ends; or referred for rehabilitation and sub-acute care and/or community care before their final discharge. A point to note is that although the diagram shows the different stages of the healthcare journey, not all patients will experience all stages of the journey. The level of care required and the coordination among careers could determine the path, as well as the duration of the journey.

As in the Australia, in most countries healthcare delivery requires coordination among several healthcare professionals. The importance of collaboration and coordination is clearly reflected in case of *task-dependency*. For example, when a specialist wishes to explore the presenting problem in greater detail, the patient is sent to a diagnostic centre for further pathology and radiology services. The specialist may also request the patient to come back for a visit a couple of days later. If the diagnostic centre does not provide the results of medical tests within a specified time then the physician may not be able to suggest any additional medication to that patient during the next visit. This kind of *task-dependency* eventually creates an interdependent network among different participating service providers, for example, between the physician and diagnostic centre. Such task dependencies demand efficient collaboration and coordination for better outcomes for the patient. As the different service providers provide different aspects of care (e.g. the diagnostic centre conducts medical tests and the physician recommends the medication to patients), they do not necessarily work together. However, there is a clear need for providers to work together while treating patients with chronic problems or patients who are admitted for acute care. When a patient with a chronic problem such as diabetes or asthma is admitted to a hospital, the treatment may require several visits by multiple specialists. During the course of treatment, they may need to change their medications depending on the patient's health condition and response to other medications. In this case, a proper collaboration among these physicians is mandatory, as they work towards the shared goal of improving the patient's condition.

Gathering appropriate information regarding the collaboration among corresponding healthcare professionals is an arduous task. In Taiwan, for example, the government of Taiwan[2] has made public health services data available to encourage researchers to conduct research in this era. For our research in this paper, we have chosen private health insurance claims data set to study collaboration that occurs among providers while treating patients who are members of a private health insurance provider.

In Australia, for instance, all physicians who treat a patient during a hospital admission send their claims to the patient's private health insurer (PHI). Typically, a claim specifies details about what service is provided to the patient, by whom the service is provided and the cost of those services. There can be several claims from several providers during the course of treating a patient for a specific hospital admission.

---

[2] http://nhird.nhri.org.tw/en/

The claims data received by PHIs are unique as both business data (e.g. cost of services) and clinical data in the form of specific procedure codes as per Commonwealth Medical Benefits Schedule (CMBS)[3] are available to them via the claim forms (Srinivasan & Arunasalam, 2013). In general, health insurance claim data sets contain a large number of claims that cover a wide variety of medical services, a broad geographic area and a long time period. In addition to utilisation statistics of different medical services and procedures, health insurance claim data set reveals information about interactions among different health service providing units (e.g. physician and hospital) during the course of providing treatment to patients.

This paper presents a social-network based framework that uses the rich claim data set as the source to analyse collaborations that occur while treating admitted patients. Figure 5 shows the framework to explore the collaboration and communication networks among healthcare professionals while treating admitted patients.

[Insert Figure 5 here]

As indicated earlier this framework is based on two network theories (i.e. Bavelas' centralisation theory (Bavelas, 1950) and Freeman's centrality theory (Freeman, 1978)). These theories explain structural changes of actors within a network and their impact on individual and group performances. In order to understand the collaboration that occurs during a treatment episode, we first extract specific data items required to construct different coordination and collaboration networks that evolve during the course of providing healthcare services to patients. For example, the networks that can be constructed from this data could include: (i) coordination network among different hospital units, and (ii) physician collaboration network. These networks can then be analysed using different social network analysis techniques such as social network centrality measures and exponential random graph models.

For research analysis purposes, only social network analysis methods and approaches have been chosen since, according to Bavelas (1950) experiment and Freeman's (1978) Centralisation theory, network positions of actors have impact on their ability to perform. In this analysis, socio-demographic characteristics (e.g. age of patient, location of the hospital and experience of the physician) of the member actors of these networks are also considered. These will ultimately enable us to explore the impact and influence of these characteristics on network formation and subsequent healthcare performance. We expect this analysis to provide both positive and negative network features of the corresponding collaboration networks. Positive network features are the properties of coordination and collaboration networks that are conducive to healthcare performance. For example, if a physician collaboration network with a higher *network density* shows better healthcare performance in terms of less hospitalisation cost then the *density* is a positive network feature for that collaboration network. In contrast, negative network features are not conducive to better healthcare performance. Finally, this framework develops predictive models using the extracted network features for estimating different healthcare outcome measures (e.g. hospitalisation cost and readmission rate) that can be utilised in any predictive health analytics tool.

---

[3] http://www.mbsonline.gov.au/

## METHODOLOGY AND APPLICATION OF THE PROPOSED RESEARCH FRAMEWORK: A CASE STUDY

In this section, we present a specific case study that explores physicians' collaboration which occurred while treating patients undergoing hip replacement surgery, covering a four year period. The data is provided by an Australian not-for-profit health insurance organisation. The data set includes three distinct categories of claims: (i) ancillary claims; (ii) medical claims; and (iii) hospital claims. Ancillary claims are submitted by providers for auxiliary services such as dental, optical, physiotherapy, dietetics, etc. Medical claims are lodged by physicians, surgeons, anaesthetists and other medical providers involved in treating the patient. Hospital claims are submitted by the hospital for all hospital services such as accommodation, theatre, ICU charges, etc. For research analysis purpose, this study considers claims data of hospital admissions only for total hip replacement (THR) patients from 85 different hospitals. In these hospitals, 2352 patients were admitted during the data collection period. In total, these patients lodged 1388 ancillary claims, 69619 medical claims and 24559 hospital claims. Table 1 shows the basic statistical details of the research data set.

[Insert Table 1 here]

Collaborations among physicians, which is termed as *Physician Collaboration Network* (PCN), emerge when they visit common hospital patients (Landon et al., 2012; Uddin, et al., 2012). Figure 6 shows the construction of a PCN and the related network and performance measures.

[Insert Figure 6 here]

The left-hand portion of Figure 6 (i.e. Construction of physician collaboration network) illustrates an example of such a PCN construction. In a hospital (say *H1*), physicians *Ph.A* and *Ph.C*, visit patient *Pa.1*; *Ph.A* and *Ph.B* visit patient *Pa.2*. This is depicted in the *patient-physician* network in the top-left corner of Figure 6. The corresponding physician collaboration network (PCN) for this *patient-physician* network is demonstrated in the bottom-left corner of Figure 6. In this PCN, there are network connections between *Ph.A* and *Ph.B* and between *Ph.A* and *Ph.C* because they visit a common patient. As people have hospital admissions for a wide range of illness and patients need to be seen by several specialists, different types of PCNs evolve. For example knee surgery patients could have a particular type of PCN, and patients with coronary diseases may have a different type of PCN. Since the research data set of this study contains health insurance claim data for THR patients from 85 different hospitals, 85 PCNs evolved during the data collection period. Figure 7 shows the construction of PCN from the research data set.

[Insert Figure 7 here]

Out of these 85 PCNs, the top-5 PCNs having higher *readmission rate* are compared with the top-5 PCNs having lower *readmission rate* using exponential random graph (ERG) model in order to explore prominence of micro-structures within these two types of PCNs.

**Social Network Analysis of Physician Collaboration**

In examining physician collaborations, we consider two social network analysis measures – *degree centralisation* and *betweenness centralisation*. Centralisation is a network-level measure whereas centrality is a node-level measure. Thus, the later one needs to be explained first before describing the former one. Centrality is an important concept in studying social networks. In conceptual term, centrality measures how central an individual is positioned in a network. *Degree centrality* is one of basic measures of network centrality. For an actor, it is the proportion of nodes that are adjacent to that actor in a network. It highlights the node with the most links to other actors in a network (S Wasserman & Faust, 2003). *Betweenness centrality* views an actor as being in a favoured position to the extent that the actor falls on the shortest paths between other pairs of actors in the network. That is, actors that occur on many shortest paths between the other pair of nodes have higher *betweenness centrality* than those they do not (Freeman, 1978). A centralisation measure quantifies the range or variability of individual actor indices that were calculated using one of the centrality measures. *Degree centralisation* is used to determine how centralised the *degree* of the set of actors is in a network. The set of *degree centralities*, which represents the collection of *degree* indices of $N$ actors in a network, can be summarised by the following equation to measure network *degree centralisation* (Freeman, et al., 1979):

$$C_D = \frac{\sum_{i=1}^{N}[C_D(n^*) - C_D(n_i)]}{[(N-1)(N-2)]}$$

Where, $\{C_D(n_i)\}$ are the *degree* indices of $N$ actors and $C_D(n^*)$ is the largest observed value in the *degree* indices. For a network, *degree centralisation* (i.e. the index $C_D$) reaches its maximum value of 1 when one actor chooses all other *(N-1)* actors and the other actors interact only with this one (i.e. the situation in a *star* graph). Similarly, the set of *betweenness centralities*, which represents the collection of *betweenness* indices of $N$ actors in a network, can be summarised by the following equation to measure network *betweenness centralisation* (Freeman, et al., 1979):

$$C_B = \frac{\sum_{i=1}^{N}[C_B^{'}(n^*) - C_B^{'}(n_i)]}{(N-1)}$$

Where, $\{C_B^{'}(n_i)\}$ are the *betweenness* indices of $N$ actors and $C_B^{'}(n^*)$ is the largest observed value in the *betweenness* indices.

We develop simple linear regression models to examine the effect of the above two SNA measures: *degree centralisation* and *betweenness centralisation* on the hospital outcome measure. The outcome measure chosen is the *hospitalisation cost*. Table 2 shows the performance of these two models.

[Insert Table 2 here]

The first model which is based on *degree centralisation* does not have any statistically significant effect on the *hospitalisation cost*. This is evident from the low R square value. On the other hand, *betweenness centralisation* has a negative effect on the *hospitalisation cost* as indicated by the second model.

From the perspective of a PCN structure, a high *betweenness centralisation* indicates that the PCN follows a *star-like* or *centralised* structure since *betweenness centralisation* reaches its highest value of 1 for a *star* network. A *star-like* or *centralised* network has few actors with higher *betweenness centrality* values. This indicates that only a small number of actors play a major collaboration role. What is interesting about this finding is that it shows that a star-like network among providers reduces the value of the outcome variable – which in this case is total *hospitalisation cost*. Therefore, this offers some interesting insights to healthcare managers and hospital administrators. Encouraging collaboration, with only one or two key people coordinating the communication (in other words establishing a *star-like* or *centralised* PCN) can help in reducing the total *hospitalisation cost*. However, a PCN with a flat network structure, where members of that PCN have almost equal network participation, appears to have high a *hospitalisation cost*. We do need to understand the influence of age as a moderating variable when it comes to hospital costs. Table 3 shows the effect of *patient age* as control (or moderating) variable.

[Insert Table 3 here]

Next we developed regression models by considering each of the network measures and its product with *patient age*. To show controlling effect, the product of network measure and *patient age* must show significant association with hospital outcome variable (i.e. *hospitalisation cost*) in the egression models (Baron & Kenny, 1986). This product shows a significant association for the second model in Table 3. That means *patient age* moderates only the relation between *betweenness centralisation* and *hospitalisation cost*. Although medical studies (e.g. Landon et al., 2013) suggest that *patient age* has an impact on patient outcomes, we notice some inconsistent outcomes. Out of the two models shown in Table 3, model 2 shows statistically significant effect while model 1 does not. This can be explained by the fact that we consider the average age of all patients in calculating *patient age* for a PCN. On the other hand, studies of present healthcare literature consider *patient age* at the individual level, not at the aggregate level as in this study.

Models of social network analysis have been utilised extensively to understand the structural dynamics of various collaboration networks (Shahadat Uddin, Jafar Hamra, et al., 2013a; S Wasserman & Faust, 2003). One of the widely used social network models is exponential random graph model which can effectively identify structural properties of network formation process (Stanley Wasserman & Pattison, 1996). Next, we will use exponential random graph models to understand the micro structures that influence the development process of the PCN.

**Exponential Random Graph (ERG) modeling of physician collaborations**

An ERG model simplifies a complex structure down to a combination of basic parameters. It can effectively identify structural properties in social networks (Snijders et al., 2006). This theory-driven modelling approach also allows us to test the significance of structural parameters in the process of the formation of a given network (Snijders, et al., 2006; S Wasserman & Faust, 2003). For instance, a given cost effective PCN may be explored using ERG model to examine what micro structures play a statistically significant role in the development process of that PCN. A commonly used sub-class of ERG models is the Markov random graph in which a possible tie from $i$ to $j$ is assumed conditionally dependent only on other possible ties involving $i$ and/or $j$ (Frank & Strauss, 1986). This sub-class of ERG models is also known as the low-order model which is utilised to explore PCNs having higher and lower readmission rates. The configurations and parameters of low-order model are shown in Figure 8.

[Insert Figure 8 here]

These parameters relate to some well-known structural regularity in the network literature and represent structural tendencies in the network (e.g. mutuality and transitivity). They were chosen because they are conceptualised as forces which drive the formation of the network itself. For example, *transitivity* is conceptualised as a force which drives the formation of the network itself (the friends of our friends are more likely to be our friends). An example of a Markov random graph model for non-directed networks, with *edge* (or density), *2-star*, *3-star* and *triangle* parameters, is given below (Robins et al., 2007):

$$\Pr(X = x) = \frac{1}{k} \exp\left\{\theta L(x) + \sigma_2 S_2(x) + \sigma_3 S_3(x) + \tau T(x)\right\}$$

In Eq.(1), *θ* is the density or edge parameter and *L(x)* refers to the number of edges in the graph *x*; $\sigma_k$ and $S_k(x)$ refer to the parameter associated with *k-star* effects and the number of *k-stars* in *x*; while *τ* and *T(x)* refer to the parameter for triangles and the number of triangles, respectively. Goodness-of-fit (GOF) measure is used to test whether a given model fits the network data. A parameter estimate in the model can be assumed to have converged if the GOF index is below 0.10 (Snijders, et al., 2006).

Using Pnet[4], five PCNs with highest *readmission rate* and five PCNs with lowest *readmission rate* were fitted with a low-order model, i.e. *2-star*, *3-star* and *triangle* model. Here, the GOF is less than 0.1, showing a good fit of the parameters of the model. Out of the three parameters, only the *triangle* parameter (subset of 3 nodes in which each node is connected to the rest 2 node) is found to have significant effect for all PCNs. The significance for each parameter is replicated by another variable, called *t-statistics*. A *t-statistics* value of ≥2 is considered to have significance effect for a given parameter. Only *triangle* parameter shows t-statistics value ≥2 for all 10 PCNs. As a part of ERG output, Pnet also produces values of *t-statistics* for each parameter of the model. The positive *triangle* parameter can be interpreted as providing evidence that the ties tend to occur in triangular structures and hence will cluster into clique-like forms. A *t-test* further reveals that there is a significant difference for the *triangle* parameter between PCNs with higher readmission rate and PCNs with lower readmission rate (t (10) = -3.05, p<0.05). The result shows that, on average, the *triangle* parameter for PCNs with higher *readmission rate* (Mean = 16.49, Standard Error = 1.47) is more positive than the parameter for PCNs with lower *readmission rate* (Mean = 5.90, Standard Error = 1.48). Therefore, the presence of a higher number of *triangle* structures in PCNs with higher *readmission rate* indicates that most actors are likely to have more connections with others. That means all network actors have almost same number of connections with others. On the other hand, PCNs with lower *readmission rate* do not have prevalence of such actors in the network structures, indicating that the possibility of the presence of some actors that have more connections within the network compared to others.

In summary, this case study illustrates the application of the proposed research framework for analysing healthcare collaboration. We have used social network analysis measures and exponential random graph models to explore physician collaborations. Further, we have made some observations on social network attributes of effective and efficient collaborations among physicians, in relation to *hospitalisation cost* and *readmission rate*.

---

[4] http://www.sna.unimelb.edu.au/pnet/pnet.html

**DISCUSSION AND CONCLUSION**

In this paper we have presented a framework based on network theories to: (i) explore the nature of collaboration among physicians while providing patient care; (ii) develop a set of network measures that provide insights into both positive and negative impact of collaboration on healthcare performance; and (iii) analyse the network values of centrality and centralisation and provide some intuitive evidence of their impact on healthcare performance as well as health outcomes for the patient.

The work presented in this paper is firmly grounded in both theory and practice as we illustrate the impact network measures in real life data from the private health insurance domain. In order to make the framework practical for health care providers, we have provided illustrative examples of physicians' collaboration while treating patients with a specific problem of hip replacement surgery. While a focussed healthcare domain example has been used, the framework itself is generic and offers a platform to study the impact of collaboration among actors involved in providing services towards achieving a common goal.

Physician collaborations have been analysed by following the proposed research framework of this study. In doing so, we considered measures of social network analysis and exponential random graph models as methodological techniques or tools, patient age as moderating factor and hospitalisation cost as the outcome measure. Based on the assumption that collaboration emerges between two physicians when they visit a common patient, PCNs are constructed in this study from the information of physicians' visits to patients during their hospitalisation periods. It is a standard professional practice that when physicians visit patients they give advice or suggestions to patients based on their health condition and previous medication history deposited in the patient *log book*. All previous advice or suggestions prescribed by any physician to a patient have been taken into consideration during any subsequent physician visit to that patient. It is noticed that the differences in performance measures (e.g. *hospitalisation cost* and *readmission rate*) among PCNs having various network structures can be explained by social network measures (i.e. *betweenness centrality*) and exponential random graph model (i.e. *triangle* parameter of the model).

The findings from the research example of this study could be utilised by hospital administrators to reflect on the nature of collaborations that occur during the course of treating a patient and encourage collaboration models that lead to better outcomes for patients. For example, a low *betweenness centralisation* in a PCN indicates a lower hospitalisation cost for the hospitalised patients, which could indirectly reflect an improved level of healthcare coordination among healthcare professionals in that PCN. Managers could encourage a practice culture where a few physicians have significantly higher number of links with their colleagues.

Although insights gained from this research have interesting applications, there are several limitations. Firstly, this study developed network structures of physician collaborations based on the presence of shared patients using administrative health insurance claim data. Although this technique has been validated (Shahadat Uddin, Liaquat Hossain, Jafar Hamra, et al., 2013), it nevertheless cannot be known what information or behaviour, if any, have an impact on the ties defined by shared patients. Secondly, although the proposed research framework can examine various collaboration networks among healthcare professionals in respect of their perceived level of desired outcomes, it cannot explain the mechanism by which these collaborations evolve over time. Finally, the possibility of unobserved confounders that could help to explain the mechanisms driving the associations that were found in this study cannot be ruled out, although patient age was used as a covariate to adjust these associations.

This research has the potential to open up several new research opportunities for healthcare researchers. Using the proposed framework of this study, researchers could explore other type of collaboration networks (e.g. collaboration networks among different hospital units) in order to find out effective and efficient collaboration structures among healthcare professionals for providing healthcare services. Another area of research is to consider qualitative measures (e.g. patient satisfaction (Reid et al., 2010)), in addition to quantitative measures, to explore the performance of healthcare collaborations. The proposed framework could be used to study physician collaboration networks for chronic patients, as well as for patients in intensive care and others requiring high level of care due to multiple comorbidities. Other social network measures (e.g. closeness centrality) and modelling approach (e.g. stochastic actor-oriented models) could be utilised for analysing and modelling physician collaboration networks for treating more complicated conditions. Finally, node-level attributes of physicians (e.g. level of education and year of experience of healthcare professional) may be considered while exploring collaboration networks.

**List of Tables**

**Table 1:** Basic statistics of the research data set

| Item | Value (Std) |
|---|---|
| Number of patients | 2352 |
| Average LoS of patients | 10.51 |
| Average age of patients | 65.02 |
| Gender distribution of patient | |
| *Female* | *1302* |
| *Male* | *1050* |
| Number of different types of claims | |
| *Hospital claim* | *24559* |
| *Medical claim* | *69619* |
| *Ancillary claim* | *1388* |

**Table 2:** Linear regression models between each of network attributes (i.e. *degree centralisation* and *betweenness centralisation*) of PCN and hospital performance measure (i.e. *hospitalisation cost*)

| Model | Dependent Variable | Independent Variable | $R^2$ value | β | Constant | Significance |
|---|---|---|---|---|---|---|
| 1 | Hospitalisation cost | Degree centralisation | 0.012 | 5906.42 | 19545.39 | 0.309 |
| 2 | Hospitalisation cost | Betweenness centralisation | 0.107 | -12384.79 | 27101.96 | 0.015 |

**Table 3:** Linear regression models for checking controlling effect of *patient age* on the relation between each of network attributes (i.e. *degree centralisation* and *betweenness centralisation*) of PCN and hospital performance measure (i.e. *hospitalisation cost*)

| Model | Dependent Variable | $R^2$ Value | Constant | Independent Variable | β | Significance |
|---|---|---|---|---|---|---|
| 1 | Hospitalisation cost | 0.102 | 20016.04 | Degrees centralisation | -26621.75 | 0.084 |
| | | | | Degree centralisation*Age | 463.16 | 0.102 |
| 2 | Hospitalisation cost | 0.227 | 27106.44 | Betweenness centralisation | -102698.68 | 0.000 |
| | | | | Betweenness centralisation*Age | 1318.36 | 0.000 |

**Key terms**

Physician collaboration network: Physicians collaborate among themselves in order to provide effective healthcare services to patients. This leads to the development of an informal collaboration network among physicians, which is termed as physician collaboration network in this study.

Social network: A social network is a set of actors (e.g. individuals or organisation) and relations (e.g. friendship, kinship, common interest, sexual relationship and financial exchange) that hold the actors together.

Social network analysis: Social network analysis is the mapping and measuring of relationships among actors in a social network, which provides both a visual and mathematical analysis of network relations among participating actors.

Degree centrality: It is one of basic measures of the network centrality. For an actor, it is the proportion of other actors that are adjacent to that actor in a network. It highlights the actor with the most links to other actors in a network.

Degree centralisation: It quantifies the range or variability in the degree centrality values of individual actors in a social network.

Betweenness centrality: It views an actor as being in a favoured position to the extent that the actor falls on the shortest paths between other pairs of actors in the network. That is, actors that occur on many shortest paths between the other pair of nodes have higher betweenness centrality than those they do not.

Betweenness centralisation: It quantifies the range or variability in the betweenness centrality values of individual actors in a social network.

Exponential random graph model: An exponential random graph model simplifies a complex structure down to a combination of basic network structure (e.g. edge). It can effectively identify structural properties in social networks.